\documentclass[letter]{aa}
\usepackage{cancel}
\usepackage{graphicx}
\usepackage{txfonts}
\usepackage{lipsum}
\usepackage{subcaption}     
\usepackage{placeins}       
\usepackage[colorlinks=true,citecolor=blue]{hyperref}

\newcommand{\nifs}{$^{56}$Ni}
\newcommand{\Nimass}{M($^{56}$Ni)}
\newcommand{\FIG}[1] {Figure~\ref{#1}}
\newcommand{\orcid}[1]{\href{https://orcid.org/#1}{\includegraphics[width=9pt]{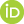}}}
\usepackage{natbib}

\begin{document}

\titlerunning{SN~2022jli hydrodynamic model of the light curve}
\authorrunning{M. Orellana et al.}
\title{SN~2022jli modeled with a $^{56}$Ni double-layer and a magnetar}
   
\author{
M.~Orellana 
\inst{\orcid{0009-0000-3624-8137},1,2}\thanks{morellana@unrn.edu.ar},
M. C. Bersten \inst{3,4,5}
\and C.~P.~Guti\'errez \inst{\orcid{0000-0003-2375-2064},6,7}}
   \offprints{Mariana Orellana}
\institute{Laboratorio de Investigación Científica en Astronomía, UNRN, Sede Andina, Mitre 630 (8400) Bariloche, Argentina
\and Consejo Nacional de Investigaciones Científicas y Técnicas (CONICET), Argentina
\and Instituto de Astrof\'isica de La Plata (IALP), CCT-CONICET-UNLP. Paseo del Bosque S/N, B1900FWA, La Plata, Argentina 
\and Facultad de Ciencias Astronómicas y Geofísicas, UNLP, Paseo del Bosque S/N 1900 La Plata, Buenos Aires, Argentina.
\and Kavli IPMU (WPI), UTIAS, The University of Tokyo, Kashiwa, Chiba 277-8583, Japan.
\and Institut d'Estudis Espacials de Catalunya (IEEC), Edifici RDIT, Campus UPC, 08860 Castelldefels (Barcelona), Spain
\and Institute of Space Sciences (ICE, CSIC), Campus UAB, Carrer de Can Magrans, s/n, E-08193 Barcelona, Spain
}
\date{Accepted 25/07/2025}
\abstract
{We study the bolometric evolution of the exceptional Type Ic Supernova (SN) 2022jli, aiming to understand the underlying mechanisms responsible for its distinctive double-peaked light curve morphology, extended timescales, and the rapid, steep decline in luminosity observed at around 270 days after the SN discovery.}
{We present a quantitative assessment of two leading models through hydrodynamic radiative simulations: two shells enriched with nickel and a combination of nickel and magnetar power.}
{We explore the parameter space of a model in which the SN is powered by radioactive decay assuming a bimodal nickel distribution. While this setup can reproduce the early light curve properties, it faces problems to explain the prominent second peak. We therefore consider a hybrid scenario with a rapidly rotating magnetar as additional energy source.}
{We find that the observed light-curve morphology can be well reproduced by a model combining a magnetar engine and a double-layer $^{56}$Ni distribution. The best-fitting case consist of a magnetar with a spin period of $P\simeq 22$~ms and a bipolar magnetic field strength of $B\simeq 5\times 10^{14}$~G and a radioactive content with total \Nimass\ of 0.15~M$_\odot$, distributed across two distinct shells within a pre-SN structure of 11~M$_\odot$. To reproduce the abrupt drop in luminosity at $\sim 270$\,d, the energy deposition from the magnetar must be rapidly and effectively switched off.}
{}

\maketitle

\section{Introduction}

Type Ic supernovae (SNe~Ic) are thought to be core-collapse supernovae (SNe) resulting from the final explosion of massive stars that have been stripped of both their hydrogen and helium envelopes before the explosion. Their spectra are defined by the absence of hydrogen and helium lines \citep[e.g.][]{Filippenko97, Gal-Yam17, Modjaz19}. 
Although the mechanism of total helium removal remains an open question in stellar evolution \citep{2020Ertl}, SNe~Ic progenitors are debated to be either very massive stars that lose their outer layers through strong stellar winds \citep{Heger03, Georgy09} or stars in binary systems that were stripped via interaction with a companion \citep[e.g.,][]{Podsiadlowski92, Nomoto95, Eldridge08}. 

The observed diversity among SNe~Ic has grown significantly in recent years, largely thanks to the advent of wide-field sky surveys.
One remarkable example is SN~2022jli. The optical light curves of this SN displayed an unusual re-brightening approximately one month after discovery as reported by \citet{2023Moore, 2024Chen, 2024Cartier}. A conspicuous second peak, comparable in luminosity to the first peak, reached maximum light around $\gtrsim$59 days post-discovery, making this an unprecedented light curve for which \cite{2024Cartier} determined the epoch of the first maximum brightness ($t_{\mathrm{max}}$) to be MJD$= 59709.6 \pm 1.2$\,days, based on a polynomial fit. This corresponds to roughly five days after its discovery on 2022 May 5 \citep{disc_report}. However, the explosion time remains poorly constrained, with the last non-detection occurring 87.5 days before the discovery \citep{2024Chen}. 

The relatively nearby distance of SN~2022jli ($\sim 23$~Mpc; \citealt{2023Moore}) enabled high-cadence follow-up through extensive photometric and spectroscopic campaigns. One of the most remarkable features of SN~2022jli is the presence of periodic fluctuations in its light curves during the decline phase. These modulations, with a period of  12.4~days and an amplitude of $\sim$1\% of the SN's peak luminosity, persist for at least $\sim$200 days \citep{2023Moore}. Spectroscopically, helium lines are weak, while hydrogen lines become visible only after the second peak. Notably, the H$\alpha$ feature begins to exhibit a periodic shift in its peak, following the light curve undulations \citep{2024Cartier}. \cite{2024Chen} suggested a tentative association with a Fermi $\gamma$-ray source, although no corresponding emission was detected in the radio or X-ray bands. \cite{2024Cartier} further extended the optical and near-infrared monitoring of SN 2022jli up to 600 days after discovery, revealing a significant near-infrared (NIR) excess from hot dust emission around $\sim$238 days.

Assuming the explosion of SN~2022jli occurred shortly before its discovery, its bolometric light curve (LC) has some resemblance to other stripped-envelope SNe such as SN~2005bf \citep{Folatelli06}, SN~2008D \citep{2008Soderberg,2008Chevalier, 2009Modjaz}, PTF11mnb \citep{Taddia18} and SN~2019cad \citep{Gutierrez2021}. These events were similarly discovered during a phase of early rise prior to the first peak of the LC, and have been modeled successfully using a double distribution of \nifs. In this context, the LC morphology of SN~2022jli provides a compelling case for testing both double \nifs\ parametric power models \citep{2peaked} and hybrid models combining radioactive decay and magnetar power source, which have been also proposed for PFT11mnb \citep{Taddia18} and SN2019cad \citep{Gutierrez2021}.

While various competing scenarios have already been discussed in the literature \citep{2023Moore, 2024Chen, 2024Cartier}, detailed hydrodynamic calculations can achieve a more robust assessment. 
\cite{2023Moore} considered several interpretations, including a combination of early shock-cooling emission from interaction with a circumstellar material (CSM) and a subsequent radioactively powered main peak. Using the MOSFiT code by \cite{2017Nicholl}, they found that explaining the duration of the early excess required a substantial CSM mass ($>3$~M$_\odot$), which is unusually high for a Type Ic SN. This would need an exotic mass-loss mechanism shortly before the explosion. Moreover, a dense CSM structure attached to the progenitor star would likely be inconsistent with the observed early bolometric LC rise.
Their modeling also implied a rather large nickel mass, M($^{56}$Ni) $= f_{\rm Ni}\, M_{\rm ej}\,= 0.234$~M$_\odot$, which add further complications. This scenario could be further explored, and also a combination between CSM and magnetar, but a detailed analysis of the CSM case is out of the focus of this work.

Accretion-powered scenarios were also discussed as possible explanations for the late-time luminosity evolution. While not ruled out, such models and mechanisms involving collision with a binary companion remain speculative and require further investigation. Particularly, the periodic light curve undulations were attributed to binarity \citep{2024Chen}, a hypothesis that has since been revised by \cite{2024King}, who proposed that SN~2022jli may mark the ultraluminous birth of a low-mass X-ray binary. 

More recently, \cite{2024Cartier} explored a scenario in which the first peak is powered by radioactive decay, modeled using Arnett's prescriptions, while a magnetar powers the second peak. Their semi-analytic one-zone model based on \cite{KB2010} reproduces the observed LC but requires the magnetar to be artificially activated $\sim 37$~days after the explosion; a significant caveat that lacks a clear physical justification. Furthermore, their dataset is temporally sparse compared to the more densely sampled observations presented by \cite{2024Chen}. We use the later observational data in this work, which include their construction of $L_{\rm pseudo}= L_{\rm BVRI}+L_{\rm NIR}$ and suitable correction to obtain $L_{\rm bol}$ \footnote{The parameters obtained with our model are subject to the assumption that the contribution outside 3750 - 25000\,\AA\ is small, as suggested by \cite{2024Chen}.}.

To quantitatively test the leading scenarios and understand the LC morphology of SNe~2022jli, we adopt the radiation hydrodynamic modeling framework developed by \cite{2peaked}, based on a one-dimensional LTE code \citep{Bersten2011}. This method enables us to distinguish between competing models involving either a double-peaked nickel distribution or a hybrid of nickel decay and magnetar power. Our approach is compatible with a range of physical configurations, including explosions in binary systems \citep[][and references]{2022Chrimes, 2024Wei} or non-standard explosion mixing-out some radioactive material \citep[e.g.][]{2021Aloy}. 
In particular, the stratified nickel structure observed in SN~2005bf-like events has been attributed to jet-like outflows during core collapse, a scenario that may also apply to SN~2022jli. 
Here we focus on reproducing the general evolution of the bolometric LC from early epochs
but we would not attempt to explain the periodic variability in SN~2022jli.

\section{Modeling the LC with two $^{56}$Ni shells}

\cite{2peaked} aimed to explain some of the diversity seen in the LC morphologies of double-peaked SNe, highlighting that an initial bolometric rise before the two peaks, such as observed in SN~2022jli, can be explained by a bimodal distribution of radioactive nickel. Their study showed that when the LC maxima are separated from each other by a long interval, typically of the order of a month, a massive progenitor is required. This ensures sufficient spatial separation of the \nifs\ components in the mass coordinate, allowing their distinct influence on the LC to manifest at different times. A nickel-poor zone between these two shells enables a marked dip between the peaks, as is the case for SN~2022jli.
That reasoning motivates our choice of a pre-SN structure with a relatively large mass.
Here we prefer a progenitor structure which we refer to as {\tt He11} and corresponds to a Zero Age Main Sequence mass of 30~M$_\odot$, evolved using the  {\sc MESA} code \citep{2011Paxton}. For our {\tt He11} model, the ejected mass is 9.55~M$_\odot$, consistent with the mass ranges estimated by \cite{2023Moore}, who inferred $M_{\rm ej}\approx 12\pm6$~M$_{\odot}$ based on the long rise to the second maximum. 

After setting the progenitor mass, the explosion energy, $E_{\rm exp}$, must be sufficiently high to impulse the massive ejecta and reproduce the observed photospheric velocities. Spectroscopic measurements at $\approx 16$ days post explosion suggest an average Fe II velocity of $v^{\mathrm{avg}}_{\ion{Fe}{II}} \simeq 8250$\,km\,s$^{-1}$\citep{2024Cartier}. We experimented with several energy values and ultimately adopted $E_{\rm exp} = 3\times 10^{51}$ erg for our simulation. This choice adequately reproduces the maximum measured velocity, though we do not intend to reproduce the entire velocity evolution. In all the calculations, we adopted a constant gamma-ray opacity of $\kappa_\gamma =0.03$ cm$^2$/g.

Following the formalism of \cite{2peaked}, we performed one-dimensional radiative transfer hydrodynamic calculations varying the fractional mass coordinates for the {\tt He11} structure with different configurations of the two \nifs\ enriched shells. The inner shell is located between $f_0\cdot M_{\rm ej}$ and $f_1\cdot M_{\rm ej}$ in terms of mass coordinate, and the outer shell is located between $f_2\cdot M_{\rm ej}$ and $f_3\cdot M_{\rm ej}$. The corresponding abundances of nickel are $X_{\rm in}$ and $X_{\rm out}$, respectively.

\begin{figure}[h!]
\centering
\includegraphics[width=\hsize]{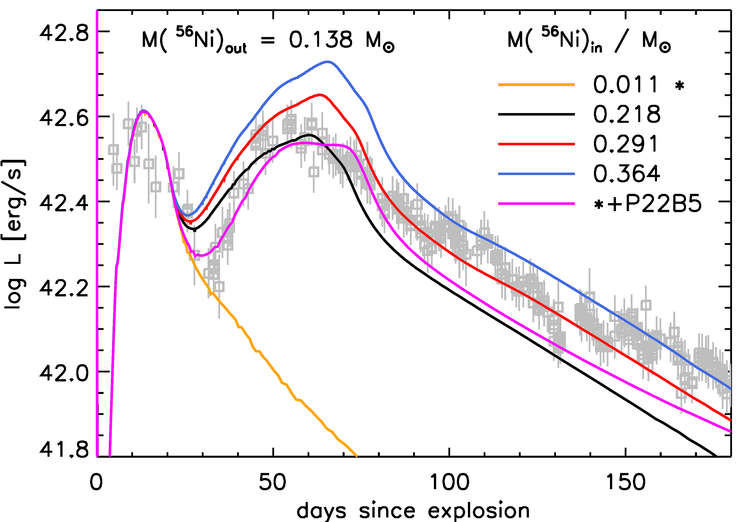}
\caption{LC comparison between the SN observations (grey markers; \citealt{2024Chen} bolometric) and the bolometric output from our double-peaked $^{56}$Ni distribution model. The inner component of the distribution is varied, while a fixed outer component of the profile is set to provide a reasonable fit for the first maximum of the LC. The model including a magnetar component is shown in magenta. In all the cases, the external enriched $^{56}$Ni shell is fixed and the progenitor is the {\tt He11} as detailed in the text.}
\label{fig1}
\end{figure}

The first LC peak occurs $\sim$ 5 days after discovery (see {Figure~\ref{fig1}})
with a bolometric peak luminosity of $L\sim 10^{42.5}$~erg\,s$^{-1}$, which can be explained by radioactive heating from the external \nifs\ shell extending up to $f_3=0.99$, i.e. nearly reaching the stellar surface. However, this outer extent is poorly constrained, as it is sensitive to the assumed explosion time. In our preferred model, the first maximum occurs around $\sim 13$ days post-explosion.
However, given the scarcity of early data (during the first ascent) and the large uncertainty associated with the timing of the explosion, the exact shape of the early LC is not entirely clear and the same applies to the parameters obtained from its modeling. Therefore, we caution against over-interpreting the shape of the first peak.

The subsequent LC decline over the following 20 days is consistent with the outer shell with the inner boundary at about $f_2\simeq0.85$ and $X_{\rm out} \simeq 0.09$. This outer component contains approximately $0.138$~M$_\odot$\ of \nifs. For a total \Nimass\ of 0.15~M$_\odot$\ (as estimated by \cite{2024Chen} and consistent with other SNe~Ic), the inner shell contains $0.011$~M$_{\odot}$ of nickel. The resulting profile is inverted, with higher \nifs\ abundance in the outer layers. However, this inversion is less extreme than in SN~2008D models found by \cite{2013Bersten}, due to the different timescales involved.
Inverted double nickel profiles can be expected in combination with jet-like outflows that might be responsible of the external placing of nickel \citep{2019Piran,2021Bugli}, or mixing instabilities \citep{2010Hammer} and see also the arguments in \cite{2peaked}.

To test whether the second LC maximum at $\sim60$~d could be powered solely by the inner \nifs\ component, we allowed the total \Nimass\ to vary freely 
between $0.138$ and $0.6$~M$_\odot$.
For a compact object (CO) a neutron star of $M_{\rm co}= 1.45$~M$_\odot$, the innermost boundary corresponds to $f_0=0.132$. After exploring different configurations, we found that a second peak timely consistent with the data can be obtained for $f_1=0.2$, and $X_{\rm in}$. 

\FIG{fig1} shows our results for the double \nifs\ distribution models contrasted with the bolometric data of SN\,2022jli. Note that these models are not fully satisfactory. If we adjust the model to match the second maximum, the late-time data ($>100$~days) is significantly underestimated (black line of Fig. 1). Conversely, if the late decline is well reproduced, the second peak appears overluminous (red and blue lines of Fig. 1). Moreover, the LC minimum brightness around $\sim30 $~d, between the two peaks, is difficult to reproduce under this scenario.  
If the nickel alone scenario were to account for the late steep decline observed at $\sim 270$~d in \cite{2024Chen} data, a strong change in the $^{56}\mathrm{Ni}$ energy leakage should be invoked. However, that hypothesis would make it difficult to explain the observed $L\sim 1.4\times 10^{40}$~erg\,s$^{-1}$ at 400~d (\citealt{2024Cartier}, not shown in our plots).

The synthetic LCs that roughly emulate the second maximum of the observed data require a non-inverted \nifs\ profile with $X_{\rm in} \sim 0.3 - 0.5$, which determine a total M($^{56}\mathrm{Ni}$)  in the range of $\approx 0.356 - 0.502$~M$_\odot$.
In comparison to more usual SNe~Ic with median M($^{56}\mathrm{Ni})$ = 0.155~M$_\odot$ \citep{Anderson2019}, 
an average of $\sim 0.22$~M$_\odot$ found by \cite{2016Lyman} and 0.16~M$_\odot$ for the sample of \cite{2016Prentice}, the aforementioned result for SN\,2022jli makes the explanation with the second peak nickel powered less plausible, therefore we explore another powering mechanism in the next Section.

\section{Magnetar as an additional power source for the second maximum} 

In order to improve the fit to the bolometric LC, we explored an alternative energy source: the spin-down of a magnetar \cite{2007Maeda}. In this scenario, the magnetar forms during the core-collapse explosion and persistently brakes, losing rotational energy which can be assumed to be deposited in the base of the SN ejecta providing an extra source of energy \citep{2010Woosley, 2017Kasen}. In our models, this power input follows the standard vacuum dipole braking index\footnote{We have found that small deviations from $n=3$ have a slight effect. Major variations of $n$ were explored by \cite{2020Orellana} and \cite{2024Omand}.}.

A more realistic approach should consider the spectral energy distribution of the magnetar and its wind (e.g. \citealt{2007Metzger, 2008Thompson}), including effects from variable opacity of the ejecta, which could lead to different regimes of propagation \citep{2010Medin}. Our treatment of the magnetar energy deposition is roughly valid at early times from the explosion \citep{2021magnetar}, but it becomes less reliable in the later ($\ge 200$\,d), post-photospheric phases.

To reproduce the second maximum of the LC of SN 2022jli, we adopted magnetar parameters of initial spin period $P\simeq 22$~ms and magnetic field strength $B\simeq 5\times 10^{14}$~G, yielding a rotational energy of $\sim 4\times 10^{49}$~erg, considerably less than the explosion energy, and a spin-down timescale of $t_p \sim 92$~days. 
The magnetar is combined with a small amount of \nifs\ retained in the innermost ejecta and the outer shell as detailed in the previous Section. The resulting LC, shown in \FIG{fig1} (magenta line), reproduces the luminosity minimum around $\sim 30$~ days more accurately than models based solely on radioactive power.

\begin{figure}[h!]
\centering
\includegraphics[width=\hsize]{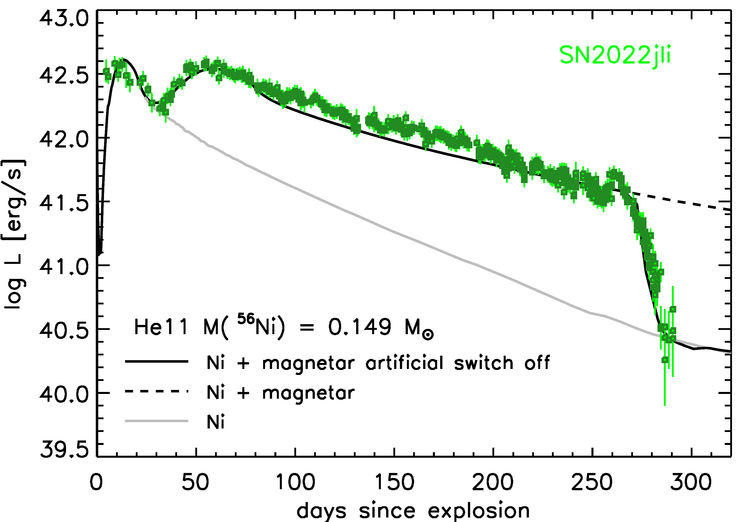}
\caption{Comparison between our preferred double-peaked \nifs\ model plus magnetar (black solid line) and the SN bolometric data from \cite{2024Chen}.}
\label{fig2}
\end{figure}

The observed LC shows a prolonged decline after the second peak, lasting until approximately $\sim 270$~days post-explosion, followed by a sudden drop in luminosity (see \FIG{fig2}). This behavior is reminiscent of the late-time evolution seen in the hydrogen-poor superluminous SN 2020wnt \citep{2022Gutierrez}, although the cause of this decline remains uncertain. 
For SN~2022jli, such a sharp drop is consistent with a sudden shutoff of the extra energy input, as mentioned by \citep{2024Chen}. Although data beyond this point are sparse, a late-time measurement at $\sim 400$~days by \cite{2024Cartier} indicates a luminosity of $L \sim 1.4\times 10^{40}$~erg~s$^{-1}$, consistent with residual power from the decay of $\approx 0.15$~M$_\odot$\ of \Nimass, similar to our inferred value. However, we note that our code does not treat properly the nebular-phase radiative processes nor the presence of dust.

While our model does not explicitly reproduce the steep decline around 270 days, this would effectively mimic a shutdown of the central engine, leaving the radioactive decay as the dominant source. 
\FIG{fig2} shows a LC in black solid line where we calibrated the onset of magnetar power suppression (i.e. the magnetar is no longer an energy source by an {\em had oc} switch off at 270 d). The transition is rapid but not instantaneous, taking about $\sim 18$~d for the LC to return to the decline rate expected from the nickel decay alone. Similar mechanisms invoking variable thermal energy injection from a magnetar have been proposed to explain other LC morphologies, such as bumps \cite{2022Moriya, 2022Chugai}. Moreover, magnetars and pulsars frequently exhibit erratic spin-down behaviour, including sudden spin-down rate variations with no substantial dependence on the spin-frequency \citep{2025Lower}. In combination with the spin, the magnetic field might change leading to a noticeable effect on the magnetar power \citep{2011Kondic, 2016Torres}.

\section{Conclusions}

Among the growing diversity of double-peaked SNe, SN~2022jli stands out as a particularly remarkable case. In this work, we have focused on modeling the overall LC as a powerful diagnostic tool to infer the physical parameters of the explosion, independently of other observational studies. Our analysis shows that the model relying solely on a double-peaked distribution of \nifs\ faces challenges in replicating the observed LC, particularly the sharp decline at late times. Moreover, it requires a total radioactive mass M($^{56}$Ni)$\gtrsim 0.35$M$_\odot$, high compared to typical core-collapse SESNe.

We therefore explored a hybrid energy-source scenario, incorporating both radioactive decay and a magnetar as energy sources. In our preferred configuration, most of the total \Nimass\ $\approx 0.15$~M$_\odot$\ is located in the outer layers of the progenitor to account for the first peak of the LC. The late luminosity reported by \cite{2024Cartier} remains consistent with this estimate. 

While the energy input from the magnetar is significantly lower than that of superluminous SNe \citep[e.g.][]{2013Inserra, 2015Wang}, the long spin-down timescale aligns well with the timing of the second peak. 
The decline from this peak is better reproduced by energy supplied by a magnetar rather than by nickel decay alone. Although the hybrid models are not without limitations, they provide a better overall fit to the observed LC, particularly the pronounced dip between the two peaks. Furthermore, assuming that magnetar power ceases to be efficiently thermalized at $\sim 270$~days, the sharp declining phase upto $\sim$290~d is well reproduced. 
Additional observational constraints during the data gap between $\sim$290\,d and $\sim$400\,d, will be valuable for ruling out or validating competing models, either through our methods or by studying spectroscopic features (\citealt{2007Maeda, 2023Omand, 2012Dessart-Ibc} and subsequent works).

Among the scenarios we explored, we conclude that a hybrid model powered by both a double-peaked \nifs\ distribution and a magnetar is currently the most plausible explanation for SN~2022jli although our model has not explained the interesting undulations shown in the LC at late epochs. 
 
\begin{acknowledgements}

This research was partially funded by UNRN PI2020 40B1039, Programa Bilateral de intercambio CSIC-iCOOP del Consejo Superior de Investigaciones Científicas de España, and the support of CONICET through project PIP 112-202001-10034.
C.P.G. acknowledges financial support from the Secretary of Universities and Research (Government of Catalonia) and by the Horizon 2020 Research and Innovation Programme of the European Union under the Marie Sk\l{}odowska-Curie and the Beatriu de Pin\'os 2021 BP 00168 program, the support from the Spanish Ministerio de Ciencia e Innovaci\'on (MCIN) and the Agencia Estatal de Investigaci\'on (AEI) 10.13039/501100011033 under the PID2023-151307NB-I00 SNNEXT project, from Centro Superior de Investigaciones Cient\'ificas (CSIC) under the PIE project 20215AT016 and the program Unidad de Excelencia Mar\'ia de Maeztu CEX2020-001058-M, and from the Departament de Recerca i Universitats de la Generalitat de Catalunya through the 2021-SGR-01270 grant. 

\end{acknowledgements}

\bibliographystyle{aa}
\bibliography{refs}

\begin{thebibliography}{54}
\expandafter\ifx\csname natexlab\endcsname\relax\def\natexlab#1{#1}\fi

\bibitem[{{Aloy} \& {Obergaulinger}(2021)}]{2021Aloy}
{Aloy}, M.~{\'A}. \& {Obergaulinger}, M. 2021, \mnras, 500, 4365

\bibitem[{{Anderson}(2019)}]{Anderson2019}
{Anderson}, J.~P. 2019, \aap, 628, A7

\bibitem[{{Bersten} {et~al.}(2011){Bersten}, {Benvenuto}, \&
  {Hamuy}}]{Bersten2011}
{Bersten}, M.~C., {Benvenuto}, O., \& {Hamuy}, M. 2011, \apj, 729, 61

\bibitem[{{Bersten} {et~al.}(2013){Bersten}, {Tanaka}, {Tominaga},
  {et~al.}}]{2013Bersten}
{Bersten}, M.~C., {Tanaka}, M., {Tominaga}, N., {et~al.} 2013, \apj, 767, 143

\bibitem[{{Bugli} {et~al.}(2021){Bugli}, {Guilet}, \&
  {Obergaulinger}}]{2021Bugli}
{Bugli}, M., {Guilet}, J., \& {Obergaulinger}, M. 2021, \mnras, 507, 443

\bibitem[{{Cartier} {et~al.}(2024){Cartier}, {Contreras}, {Stritzinger},
  {Hamuy}, {Ruiz-Lapuente}, {Prieto}, {Anderson}, {Cikota}, \&
  {Gerlach}}]{2024Cartier}
{Cartier}, R., {Contreras}, C., {Stritzinger}, M., {et~al.} 2024,
  arXiv:2410.21381

\bibitem[{{Chen} {et~al.}(2024){Chen}, {Gal-Yam}, {Sollerman}, {Schulze},
  {Post}, {Liu}, {Ofek}, {Das}, {Fremling}, {Horesh}, {Katz}, {Kushnir},
  {Kasliwal}, {Kulkarni}, {Liu}, {Liu}, {Miller}, {Rose}, {Waxman}, {Yang},
  {Yao}, {Zackay}, {Bellm}, {Dekany}, {Drake}, {Fang}, {Fynbo}, {Groom},
  {Helou}, {Irani}, {Jegou du Laz}, {Liu}, {Mazzali}, {Neill}, {Qin}, {Riddle},
  {Sharon}, {Strotjohann}, {Wold}, \& {Yan}}]{2024Chen}
{Chen}, P., {Gal-Yam}, A., {Sollerman}, J., {et~al.} 2024, \nat, 625, 253

\bibitem[{{Chevalier} \& {Fransson}(2008)}]{2008Chevalier}
{Chevalier}, R.~A. \& {Fransson}, C. 2008, \apjl, 683, L135

\bibitem[{{Chrimes} {et~al.}(2022){Chrimes}, {Levan}, {Fruchter}, {Groot},
  {Jonker}, {Kouveliotou}, {Lyman}, {Stanway}, {Tanvir}, \&
  {Wiersema}}]{2022Chrimes}
{Chrimes}, A.~A., {Levan}, A.~J., {Fruchter}, A.~S., {et~al.} 2022, \mnras,
  513, 3550

\bibitem[{{Chugai} \& {Utrobin}(2022)}]{2022Chugai}
{Chugai}, N.~N. \& {Utrobin}, V.~P. 2022, \mnras, 512, L71

\bibitem[{{Dessart} {et~al.}(2012){Dessart}, {Hillier}, {Li}, \&
  {Woosley}}]{2012Dessart-Ibc}
{Dessart}, L., {Hillier}, D.~J., {Li}, C., \& {Woosley}, S. 2012, \mnras, 424,
  2139

\bibitem[{{Eldridge} {et~al.}(2008){Eldridge}, {Izzard}, \&
  {Tout}}]{Eldridge08}
{Eldridge}, J.~J., {Izzard}, R.~G., \& {Tout}, C.~A. 2008, \mnras, 384, 1109

\bibitem[{{Ertl} {et~al.}(2020){Ertl}, {Woosley}, {Sukhbold}, \&
  {Janka}}]{2020Ertl}
{Ertl}, T., {Woosley}, S.~E., {Sukhbold}, T., \& {Janka}, H.~T. 2020, \apj,
  890, 51

\bibitem[{{Filippenko}(1997)}]{Filippenko97}
{Filippenko}, A.~V. 1997, \araa, 35, 309

\bibitem[{{Folatelli} {et~al.}(2006){Folatelli}, {Contreras}, {Phillips},
  {Woosley}, {Blinnikov}, {Morrell}, {Suntzeff}, {Lee}, {Hamuy},
  {Gonz{\'a}lez}, {Krzeminski}, {Roth}, {Li}, {Filippenko}, {Foley},
  {Freedman}, {Madore}, {Persson}, {Murphy}, {Boissier}, {Galaz},
  {Gonz{\'a}lez}, {McCarthy}, {McWilliam}, \& {Pych}}]{Folatelli06}
{Folatelli}, G., {Contreras}, C., {Phillips}, M.~M., {et~al.} 2006, \apj, 641,
  1039

\bibitem[{{Gal-Yam}(2017)}]{Gal-Yam17}
{Gal-Yam}, A. 2017, {Observational and Physical Classification of Supernovae},
  ed. A.~W. {Alsabti} \& P.~{Murdin}, 195

\bibitem[{{Georgy} {et~al.}(2009){Georgy}, {Meynet}, {et~al.}}]{Georgy09}
{Georgy}, C., {Meynet}, G., {et~al.} 2009, \aap, 502, 611

\bibitem[{{Guti{\'e}rrez} {et~al.}(2021){Guti{\'e}rrez}, {Bersten}, {Orellana},
  {Pastorello}, {Ertini}, {Folatelli}, {Pignata}, {Anderson}, {Smartt},
  {Sullivan}, {Pursiainen}, {Inserra}, {Elias-Rosa}, {Fraser}, {Kankare},
  {Moran}, {Reguitti}, {Reynolds}, {Stritzinger}, {Burke}, {Frohmaier},
  {Galbany}, {Hiramatsu}, {Howell}, {Kuncarayakti}, {Mattila},
  {M{\"u}ller-Bravo}, {Pellegrino}, \& {Smith}}]{Gutierrez2021}
{Guti{\'e}rrez}, C.~P., {Bersten}, M.~C., {Orellana}, M., {et~al.} 2021,
  \mnras, 504, 4907

\bibitem[{{Guti{\'e}rrez} {et~al.}(2022){Guti{\'e}rrez}, {Pastorello},
  {Bersten}, {Benetti}, {Orellana}, {Fiore}, {Karamehmetoglu}, {Kravtsov},
  {Reguitti}, {Reynolds}, {Valerin}, {Mazzali}, {Sullivan}, {Cai},
  {Elias-Rosa}, {Fraser}, {Hsiao}, {Kankare}, {Kotak}, {Kuncarayakti}, {Li},
  {Mattila}, {Mo}, {Moran}, {Ochner}, {Shahbandeh}, {Tomasella}, {Wang}, {Yan},
  {Zhang}, {Zhang}, \& {Stritzinger}}]{2022Gutierrez}
{Guti{\'e}rrez}, C.~P., {Pastorello}, A., {Bersten}, M., {et~al.} 2022, \mnras,
  517, 2056

\bibitem[{{Hammer} {et~al.}(2010){Hammer}, {Janka}, \&
  {M{\"u}ller}}]{2010Hammer}
{Hammer}, N.~J., {Janka}, H.~T., \& {M{\"u}ller}, E. 2010, \apj, 714, 1371

\bibitem[{{Heger} {et~al.}(2003){Heger}, {Fryer}, {Woosley}, {Langer}, \&
  {Hartmann}}]{Heger03}
{Heger}, A., {Fryer}, C.~L., {Woosley}, S.~E., {Langer}, N., \& {Hartmann},
  D.~H. 2003, \apj, 591, 288

\bibitem[{{Inserra} {et~al.}(2013){Inserra}, {Smartt}, {Jerkstrand}, {Valenti},
  {Fraser}, {Wright}, {Smith}, {Chen}, {Kotak}, {Pastorello}, {Nicholl},
  {Bresolin}, {Kudritzki}, {Benetti}, {Botticella}, {Burgett}, {Chambers},
  {Ergon}, {Flewelling}, {Fynbo}, {Geier}, {Hodapp}, {Howell}, {Huber},
  {Kaiser}, {Leloudas}, {Magill}, {Magnier}, {McCrum}, {Metcalfe}, {Price},
  {Rest}, {Sollerman}, {Sweeney}, {Taddia}, {Taubenberger}, {Tonry},
  {Wainscoat}, {Waters}, \& {Young}}]{2013Inserra}
{Inserra}, C., {Smartt}, S.~J., {Jerkstrand}, A., {et~al.} 2013, \apj, 770, 128

\bibitem[{{Kasen}(2017)}]{2017Kasen}
{Kasen}, D. 2017, {Unusual Supernovae and Alternative Power Sources}, ed. A.~W.
  {Alsabti} \& P.~{Murdin}, 939

\bibitem[{{Kasen} \& {Bildsten}(2010)}]{KB2010}
{Kasen}, D. \& {Bildsten}, L. 2010, \apj, 717, 245

\bibitem[{{King} \& {Lasota}(2024)}]{2024King}
{King}, A. \& {Lasota}, J.-P. 2024, \aap, 682, L22

\bibitem[{{Kondi{\'c}} {et~al.}(2011){Kondi{\'c}}, {R{\"u}diger}, \&
  {Hollerbach}}]{2011Kondic}
{Kondi{\'c}}, T., {R{\"u}diger}, G., \& {Hollerbach}, R. 2011, \aap, 535, L2

\bibitem[{{Lower} {et~al.}(2025){Lower}, {Karastergiou}, {Johnston}, {Brook},
  {Dai}, {Kerr}, {Manchester}, {Oswald}, {Shannon}, {Sobey}, \&
  {Weltevrede}}]{2025Lower}
{Lower}, M.~E., {Karastergiou}, A., {Johnston}, S., {et~al.} 2025, \mnras, 538,
  3104

\bibitem[{{Lyman} {et~al.}(2016){Lyman}, {Bersier}, {James}, {Mazzali},
  {Eldridge}, {Fraser}, \& {Pian}}]{2016Lyman}
{Lyman}, J.~D., {Bersier}, D., {James}, P.~A., {et~al.} 2016, \mnras, 457, 328

\bibitem[{{Maeda} {et~al.}(2007){Maeda}, {Tanaka}, {Nomoto}, {Tominaga},
  {Kawabata}, {Mazzali}, {Umeda}, {Suzuki}, \& {Hattori}}]{2007Maeda}
{Maeda}, K., {Tanaka}, M., {Nomoto}, K., {et~al.} 2007, \apj, 666, 1069

\bibitem[{{Medin} \& {Lai}(2010)}]{2010Medin}
{Medin}, Z. \& {Lai}, D. 2010, \mnras, 406, 1379

\bibitem[{{Metzger} {et~al.}(2007){Metzger}, {Thompson}, \&
  {Quataert}}]{2007Metzger}
{Metzger}, B.~D., {Thompson}, T.~A., \& {Quataert}, E. 2007, \apj, 659, 561

\bibitem[{{Modjaz} {et~al.}(2019){Modjaz}, {Guti{\'e}rrez}, \&
  {Arcavi}}]{Modjaz19}
{Modjaz}, M., {Guti{\'e}rrez}, C.~P., \& {Arcavi}, I. 2019, Nature Astronomy,
  3, 717

\bibitem[{{Modjaz} {et~al.}(2009){Modjaz}, {Li}, {Butler}, {Chornock},
  {Perley}, {Blondin}, {Bloom}, {Filippenko}, {Kirshner}, {Kocevski},
  {Poznanski}, {Hicken}, {Foley}, {Stringfellow}, {Berlind}, {Barrado y
  Navascues}, {Blake}, {Bouy}, {Brown}, {Challis}, {Chen}, {de Vries},
  {Dufour}, {Falco}, {Friedman}, {Ganeshalingam}, {Garnavich}, {Holden},
  {Illingworth}, {Lee}, {Liebert}, {Marion}, {Olivier}, {Prochaska},
  {Silverman}, {Smith}, {Starr}, {Steele}, {Stockton}, {Williams}, \&
  {Wood-Vasey}}]{2009Modjaz}
{Modjaz}, M., {Li}, W., {Butler}, N., {et~al.} 2009, \apj, 702, 226

\bibitem[{{Monard}(2022)}]{disc_report}
{Monard}, L. 2022, Transient Name Server Discovery Report, 2022-1198, 1

\bibitem[{{Moore} {et~al.}(2023){Moore}, {Smartt}, {Nicholl}, {Srivastav},
  {Stevance}, {Jess}, {Grant}, {Fulton}, {Rhodes}, {Sim}, {Hirai},
  {Podsiadlowski}, {Anderson}, {Ashall}, {Bate}, {Fender}, {Guti{\'e}rrez},
  {Howell}, {Huber}, {Inserra}, {Leloudas}, {Monard}, {M{\"u}ller-Bravo},
  {Shappee}, {Smith}, {Terreran}, {Tonry}, {Tucker}, {Young}, {Aamer}, {Chen},
  {Ragosta}, {Galbany}, {Gromadzki}, {Harvey}, {Hoeflich}, {McCully},
  {Newsome}, {Gonzalez}, {Pellegrino}, {Ramsden}, {P{\'e}rez-Torres}, {Ridley},
  {Sheng}, \& {Weston}}]{2023Moore}
{Moore}, T., {Smartt}, S.~J., {Nicholl}, M., {et~al.} 2023, \apjl, 956, L31

\bibitem[{{Moriya} {et~al.}(2022){Moriya}, {Murase}, {et~al.}}]{2022Moriya}
{Moriya}, T.~J., {Murase}, K., {et~al.} 2022, \mnras, 513, 6210

\bibitem[{{Nicholl} {et~al.}(2017){Nicholl}, {Guillochon}, \&
  {Berger}}]{2017Nicholl}
{Nicholl}, M., {Guillochon}, J., \& {Berger}, E. 2017, \apj, 850, 55

\bibitem[{{Nomoto} {et~al.}(1995){Nomoto}, {Iwamoto}, \& {Suzuki}}]{Nomoto95}
{Nomoto}, K.~I., {Iwamoto}, K., \& {Suzuki}, T. 1995, \physrep, 256, 173

\bibitem[{{Omand} \& {Jerkstrand}(2023)}]{2023Omand}
{Omand}, C.~M.~B. \& {Jerkstrand}, A. 2023, \aap, 673, A107

\bibitem[{{Omand} \& {Sarin}(2024)}]{2024Omand}
{Omand}, C. M.~B. \& {Sarin}, N. 2024, \mnras, 527, 6455

\bibitem[{{Orellana} \& {Bersten}(2020)}]{2020Orellana}
{Orellana}, M. \& {Bersten}, M.~C. 2020, BAAA, 61B, 63

\bibitem[{{Orellana} \& {Bersten}(2022)}]{2peaked}
{Orellana}, M. \& {Bersten}, M.~C. 2022, \aap, 667, A92

\bibitem[{{Paxton} {et~al.}(2011){Paxton}, {Bildsten}, {Dotter}, {Herwig},
  {Lesaffre}, \& {Timmes}}]{2011Paxton}
{Paxton}, B., {Bildsten}, L., {Dotter}, A., {et~al.} 2011, \apjs, 192, 3

\bibitem[{{Piran} {et~al.}(2019){Piran}, {Nakar}, {Mazzali}, \&
  {Pian}}]{2019Piran}
{Piran}, T., {Nakar}, E., {Mazzali}, P., \& {Pian}, E. 2019, \apjl, 871, L25

\bibitem[{{Podsiadlowski} {et~al.}(1992){Podsiadlowski}, {Joss}, \&
  {Hsu}}]{Podsiadlowski92}
{Podsiadlowski}, P., {Joss}, P.~C., \& {Hsu}, J.~J.~L. 1992, \apj, 391, 246

\bibitem[{{Prentice} {et~al.}(2016){Prentice}, {Mazzali}, {Pian}, {Gal-Yam},
  {Kulkarni}, {Rubin}, {Corsi}, {Fremling}, {Sollerman}, {Yaron}, {Arcavi},
  {Zheng}, {Kasliwal}, {Filippenko}, {Cenko}, {Cao}, \&
  {Nugent}}]{2016Prentice}
{Prentice}, S.~J., {Mazzali}, P.~A., {Pian}, E., {et~al.} 2016, \mnras, 458,
  2973

\bibitem[{{Soderberg} {et~al.}(2008){Soderberg}, {Berger}, {Page}, {Schady},
  {Parrent}, {Pooley}, {Wang}, {Ofek}, {Cucchiara}, {Rau}, {Waxman}, {Simon},
  {Bock}, {Milne}, {Page}, {Barentine}, {Barthelmy}, {Beardmore}, {Bietenholz},
  {Brown}, {Burrows}, {Burrows}, {Byrngelson}, {Cenko}, {Chandra}, {Cummings},
  {Fox}, {Gal-Yam}, {Gehrels}, {Immler}, {Kasliwal}, {Kong}, {Krimm},
  {Kulkarni}, {Maccarone}, {M{\'e}sz{\'a}ros}, {Nakar}, {O'Brien}, {Overzier},
  {de Pasquale}, {Racusin}, {Rea}, \& {York}}]{2008Soderberg}
{Soderberg}, A.~M., {Berger}, E., {Page}, K.~L., {et~al.} 2008, \nat, 453, 469

\bibitem[{{Taddia} {et~al.}(2018){Taddia}, {Sollerman}, {Fremling},
  {Karamehmetoglu}, {Quimby}, {Gal-Yam}, {Yaron}, {Kasliwal}, {Kulkarni},
  {Nugent}, {Smadja}, \& {Tao}}]{Taddia18}
{Taddia}, F., {Sollerman}, J., {Fremling}, C., {et~al.} 2018, \aap, 609, A106

\bibitem[{{Thompson}(2008)}]{2008Thompson}
{Thompson}, C. 2008, \apj, 688, 499

\bibitem[{{Torres-Forn{\'e}} {et~al.}(2016){Torres-Forn{\'e}},
  {Cerd{\'a}-Dur{\'a}n}, {et~al.}}]{2016Torres}
{Torres-Forn{\'e}}, A., {Cerd{\'a}-Dur{\'a}n}, P., {et~al.} 2016, \mnras, 456,
  3813

\bibitem[{{Vurm} \& {Metzger}(2021)}]{2021magnetar}
{Vurm}, I. \& {Metzger}, B.~D. 2021, \apj, 917, 77

\bibitem[{{Wang} {et~al.}(2015){Wang}, {Wang}, {Dai}, \& {Wu}}]{2015Wang}
{Wang}, S.~Q., {Wang}, L.~J., {Dai}, Z.~G., \& {Wu}, X.~F. 2015, \apj, 799, 107

\bibitem[{{Wei} {et~al.}(2024){Wei}, {Yang}, {Wei}, \& {Dai}}]{2024Wei}
{Wei}, Y.-J., {Yang}, Y.-P., {Wei}, D.-M., \& {Dai}, Z.-G. 2024, \aap, 688,
  A114

\bibitem[{{Woosley}(2010)}]{2010Woosley}
{Woosley}, S.~E. 2010, \apjl, 719, L204

\end{thebibliography}

\begin{appendix}
\section{Interaction with circumstellar material}

To further investigate the origin of SN\,2022jli, and given that circumstellar material (CSM) interaction is commonly invoked for some types of core-collapse SNe, we explored the potential contribution from CSM interaction using our code, assuming a stationary wind configuration. For SNe with the explosion date well established, these parameters can be constrained through early-time data modeling. For SN\,2022jli, in accordance with the rest of the work, we assumed the explosion coincides with the discovery date.

Although we did not conduct an thorough parameter space exploration, we performed a set of calculations, including CSM, to estimate the mass and extent required to reproduce the first peak. We assumed a CSM expanding at a constant velocity of 115~km~s$^{-1}$. For our {\tt He11} progenitor model, we find that approximately 2\,M$_\odot$ of external material is needed to reach the luminosity level of the first peak and to reproduce an evolution roughly compatible with the first 30 days of the light curve. This mass of CSM would extend up to 250 R$_\odot$, i.e. about 50 times the stellar radius. We find that our estimates are broadly consistent with the results of \cite{2023Moore}.

While a more detailed exploration of this scenario is beyond the scope of this work, a combined model involving both CSM interaction and magnetar input remains an interesting possibility that justifies further investigation. Importantly, if a specific CSM configuration is adopted, the magnetar parameters required to power the second peak would need to be recalibrated, since the total ejected mass would increase in a non-negligible way with respect to the original model.

\end{appendix}
\end{document}